\RequirePackage{silence}
\documentclass[12pt]{article}

\usepackage[top=2.0cm,bottom=2.0cm,left=2.54cm,right=2.54cm]{geometry}
\usepackage{
amsmath,
amssymb,
graphicx,subfigure,caption,
comment,
booktabs,
cite,
mathrsfs 
}
\usepackage{bm}
\usepackage{color}  
\usepackage{extarrows}
\usepackage{appendix}     
\usepackage{amsthm}
\usepackage{extarrows}

\usepackage{xcolor}
\usepackage[normalem]{ulem}
\definecolor{delcolor}{rgb}{0.8,0,0}
\renewcommand{\sout}{\bgroup\markoverwith{\textcolor{delcolor}{\rule[0.5ex]{2pt}{0.4pt}}}\ULon}

\newtheorem{theorem}{Theorem}
\newtheorem{lemma}{Lemma}
\newtheorem{corollary}{Corollary}

\usepackage[hyperindex=true,
          pdfstartview=FitH,
          bookmarksnumbered=true,
          bookmarksopen=true,
          citecolor=blue,
          linkcolor=blue,
          colorlinks=true,
          pdfborder=001,
          unicode]{hyperref}

\usepackage{orcidlink}

\allowdisplaybreaks

\newcommand{\D}{{\rm d}}
\newcommand{\ie}{\emph{i.e. }}

\newcommand{\vol}{\mathrm{Vol}}

\newcommand{\ind}[2]{^{#1}{}_{#2}}

\newcommand{\re}{{\rm e}}

\title{Geometric formulation for the relativistic \\
kinetic theory of photons}
\author{
Yifan Cai${}^1$\thanks{{\em email}: \href{mailto:caiyifan@stumail.nwu.edu.cn}
{caiyifan@stumail.nwu.edu.cn}},~
Long Cui${}^1$\thanks{{\em email}: \href{mailto:cuilong@mail.nankai.edu.cn}
{cuilong@mail.nankai.edu.cn}},~
Bin Wu ${}^{1,3,4,5}$\thanks{{Corresponding author, \em email}: 
\href{mailto:binwu@nwu.edu.cn}{binwu@nwu.edu.cn}},~
and Liu Zhao${}^2$\thanks{{Corresponding author, \em email}: 
\href{mailto:lzhao@nankai.edu.cn}{lzhao@nankai.edu.cn}}\\
\small ${}^1$School of Physics, Northwest University, Xi'an\ 710127, China\\
\small ${}^2$School of Physics, Nankai University, Tianjin 300071, China\\
\small ${}^{3}$Shaanxi Key Laboratory for Theoretical Physics Frontiers, Xi'an 710127, China\\
\small ${}^{4}$Peng Huanwu Center for Fundamental Theory, Xi'an 710127, China\\
\small ${}^{5}$Fundamental Discipline Research Center for Quantum Science and technology of\\
\small Shaanxi Province, Xi'an 710127, China
}

\begin{document}
\date{}
\maketitle

\begin{abstract}
We provide a geometric foundation for the kinetic theory of photons. 
Although the induced metric $\hat h$ on the light cone bundle $\Gamma_0^+$ is 
degenerate, which causes the corresponding volume element to vanish, 
we can still use a method similar to the Hodge dual to construct a volume 
element $\vol_{\Gamma_0^+}$ for the light cone bundle. Based on this geometric 
structure $(\Gamma^+_0,\vol_{\Gamma_0^+},\hat h)$, we establish the fully 
covariant Boltzmann equation for photons. 
More importantly, the volume element and the induced metric are linked in a nontrivial way, which allows the physical distributions to be defined consistently.
This yields the corresponding hydrodynamic quantities and their divergences, which take the same form as in the case of massive particles.
\end{abstract}
\section{Introduction}
\label{sec:intro}
In large-scale gravitational systems, quantum effects of matter are typically neglected. Instead, the contributions of matter to the gravitational field are coarse-grained into its energy momentum tensor $T^{\mu\nu}$, which encapsulates the statistical behavior of classical particles. This necessitates a statistical theory in curved spacetime. The relativistic kinetic theory for massive particles has a well-established history. Israel \cite{israel1963relativistic} derived the fully covariant relativistic Boltzmann equation in the 1960s, and Stewart \cite{stewart1971} and Ehlers \cite{ehlers1973survey} extended the theory to curved spacetime in the following decade. This theory has since become a cornerstone for nonequilibrium thermodynamics and hydrodynamics, such as transient relativistic thermodynamics \cite{stewart1977transient,israel1979transient}, divergence-type fluids \cite{reula1997causal}, and transport theory \cite{Lindquist:1966igj,Kremer2012}. Its applications span diverse fields, including stability theory \cite{banach1994,Fajman:2017gaz}, quark-gluon plasma \cite{Heinz:1983nx,Elze:1989un,Arnold:2007pg}, black hole accretion \cite{Rioseco:2016jwc,Li:2023qtd}, and cosmology \cite{Weinberg:2003ur,Agon:2011mz,vereshchagin2017relativistic}.

A key challenge in establishing the covariant Boltzmann equation arises from the fact that, in relativistic theories, density quantities typically depend on spacetime decomposition. Stewart and Ehlers \cite{stewart1971,ehlers1973survey} resolved this issue by connecting the state density with the weight value of geodesic trajectories. Their crucial idea is to first construct a volume element in the space of geodesic phase trajectories. A one-to-one mapping between this trajectory space and an arbitrary Cauchy surface within the mass shell bundle $\Gamma_m := \{(x,p) \in T\mathcal{M} \mid p^\mu p_\mu = -m^2\}$ then extends this volume element to the Cauchy surface, which represents the microstate space of particles. While this construction is mathematically rigorous, a fully geometric formulation remained desirable, as it would provide a more intuitive foundation and facilitate generalizations.

Sarbach \emph{et al.} \cite{sarbach2013rel,sarbach2014geo} further developed the relativistic kinetic theory by introducing the Sasaki metric $\hat g$ \cite{sasaki1958differential} on the tangent bundle. This metric automatically induces a metric $\hat h$ and a volume element on the mass shell bundle, and naturally describes the dependence of the state density on spacetime decomposition through the inner product between the density current and the unit normal vector of the microstate space. This geometric framework, however, relies critically on the nondegeneracy of the induced metric $\hat h$. For massless particles, $\hat h$ becomes degenerate, rendering the volume element on the future light cone bundle $\Gamma_0^+ := \{(x,k) \in T\mathcal{M} \mid k^\mu k_\mu = 0,\; -Z_\mu k^\mu > 0\}$ ill-defined. This degeneracy is the central obstacle we address in this paper.

In recent years, black hole shadows and images have attracted extensive attention as probes of gravitational dynamics in strong-field regimes~\cite{Akiyama_2019,Akiyama_2022}. These observables involve two distinct physical processes, photon emission from accretion disks and their subsequent propagation through curved spacetime. The former is typically modeled by GRMHD~\cite{Zhou:2025moa}, while the latter is computed by ray tracing~\cite{Cunha:2018acu,Gralla:2019xty}. However, ray tracing alone is insufficient, as it assumes a fixed source distribution rather than solving it dynamically. Since accretion disks are typically sufficiently rarefied, relativistic kinetic theory can provide a unified first-principles framework for describing both processes. In this paper, we address the geometric foundation of such a kinetic theory, specifically the degeneracy of the induced metric on the light cone bundle. The full kinetic description, including scattering integrals and the H theorem, will be presented in future work. Together, this work and its planned extensions are intended to provide a more refined kinetic description of the accretion flow, which may serve as a useful complement to GRMHD simulations and as a basis for more detailed predictions of black hole shadows.

To resolve the degeneracy issue, we employ a construction analogous to the Hodge dual to define a nonzero volume element $\vol_{\Gamma_0^+}$ on $\Gamma_0^+$. This volume element is not compatible with the degenerate induced metric $\hat h$; nonetheless, it allows us to recover the kinetic description in a form similar to the massive-particle case~\cite{sarbach2013rel,sarbach2014geo}. Our strategy is twofold. First, we formulate a tensorial Boltzmann equation that is independent of any geometric structure. Second, we scalarize this equation using the volume element $\vol_{\Gamma_0^+}$. Statistical quantities, such as the density current, are then defined with the aid of $\hat h$.

The paper is organized as follows: In Sec.~\ref{sec:light-cone-bundle}, 
we construct the geometric structure of the light cone bundle, including 
the volume element and the induced metric, and analyze their properties. 
In Sec.~\ref{sec:boltzmann}, we first utilize the approach provided in 
Appendix.~\ref{app:distribution-on-manifold} to establish a tensorial 
Boltzmann equation, and then derive its scalarization under the volume element 
introduced in Sec.~\ref{sec:light-cone-bundle}. With the induced metric on 
the light cone bundle, the state density distribution can be described as the 
inner product between the density current and the unit normal vector of 
the space of microstates. By integration over the momentum space, the corresponding 
hydrodynamic quantities are provided in Sec.~\ref{sec:current}, and the corresponding 
covariant divergences are also analyzed. Finally, we conclude our work and discuss 
possible applications of this theory in Sec.~\ref{sec:conclusion}.

Throughout this paper, we denote the $(d+1)$-dimensional spacetime manifold by $\mathcal{M}$ and adopt the Lorentzian signature $(-,+,\cdots,+)$ for the metric $g_{\mu\nu}(x)$. We consider a family of observers whose worldlines cover the spacetime. Their velocity field, denoted by $Z^\mu$, is assumed to be a future-directed unit timelike vector field.

\section{Light cone bundle}
\label{sec:light-cone-bundle}

The action for a relativistic particle of mass $m$ can be written as
\begin{align}\label{eq:action}
    S[x(\lambda),e(\lambda)]=\int\D\lambda\left[\frac{1}{2e}\dot{x}^2-\frac{e}{2}m^2\right],
\end{align}
where $\dot{x}^2:=g_{\mu\nu}\dot{x}^\mu\dot{x}^\nu$. The auxiliary worldline field $e(\lambda)$ ensures the reparametrization invariance of the action. In fact, $S$ is invariant under an arbitrary reparametrization
\begin{align}
    x^\mu(\lambda)\mapsto x^\mu(\alpha(\lambda)),\quad  e(\lambda)\mapsto \alpha'(\lambda) e(\alpha(\lambda)),
\end{align}
where $\alpha:\mathbb{R}\rightarrow\mathbb{R}$ is an arbitrary smooth, monotonically increasing function. The equations of motion for $x^\mu$ and $e$ are, respectively,
\begin{align}
    \dot{x}^\nu \nabla_\nu \dot{x}_\mu=\dot{e} e^{-1}\dot{x}_\mu,\qquad \dot{x}^2+e^2 m^2=0.
\end{align}
A suitable gauge condition\footnote{For massive particles, the more conventional gauge choice is $e=m^{-1}$; here, however, we adopt a condition that applies to both massive and massless particles.} that fixes the reparametrization freedom is $e=1$. Under this condition, the canonical momentum coincides with the velocity:
\begin{align}
    k_\mu=\frac{\partial L}{\partial \dot{x}^\mu}=e^{-1}\dot{x}_\mu=\dot{x}_\mu.
\end{align}
Moreover, under this gauge choice, the right-hand side of the equation of motion for $x^\mu$ vanishes. Consequently, an affine transformation of a solution, $\lambda\mapsto a\lambda+b$, does not change the equation of motion for $x^\mu$. For a massive particle, however, such a transformation does not preserve the equation of motion for $e$, namely the constraint $\dot{x}^2+m^2=0$ when $e=1$. Therefore, even if we reparametrize a solution, the original parameter can still be recovered from the equations of motion, up to the remaining translation of the parameter origin. For a massive particle, the choice of parameter is thus merely a gauge freedom and carries no physical information.

For a massless particle, an affine transformation of the parameter does not change the equation of motion for $e$. Consequently, the original parametrized solution cannot be recovered from the reparametrized solution using the equations of motion alone. This does not imply that $e=1$ is an improper gauge choice. Under an affine rescaling, the momentum transforms as $k^{\mu}\mapsto a k^\mu$. The energy (or frequency) of a photon measured by an observer 
$Z^\mu$ is
\begin{align}\label{eq:energy-photon}
\omega = -Z_\mu k^\mu,
\end{align}
and therefore transforms under the same rescaling. The residual affine freedom is thus physically relevant because it determines the energy scale of the photon; it is not a remaining gauge redundancy. If no dynamical variables other than $x^\mu$ are introduced, the affine parametrization is required to encode this physical degree of freedom. Accordingly, the present work does not pursue a reparametrization-invariant formulation. This is the principal distinction between our approach and that of Ref.~\cite{gunneberg2026vlasov}.

Throughout this work, we adopt the gauge choice $e=1$. Consequently, $k^\mu$ represents both the photon momentum and the tangent vector to its worldline with respect to the affine parameter $\lambda$. The photon is governed by the geodesic equation and the null constraint,
\begin{align}
    k^\nu\nabla_\nu k^\mu=0,\qquad k^\mu k_\mu=0.
\end{align}
Although a photon microstate is specified by $(x^\mu,k^\mu)$, the null constraint implies that the physically admissible microstates do not fill the entire tangent bundle $T\mathcal{M}$, whose dimension is $2d+2$. We therefore define the future light cone bundle as the $(2d+1)$-dimensional submanifold
\begin{align}
\Gamma^+_0=\{(x,k)\in T\mathcal M\mid \mathcal{H}(x,k)=0,\ -Z_\mu k^\mu >0\},
\end{align}
which we henceforth refer to simply as the light cone bundle. Here, $\mathcal{H}$ is the Hamiltonian corresponding to the action \eqref{eq:action} under the gauge choice $e=1$, given by
\begin{align}
    \mathcal{H}(x,k):=\frac{1}{2}g_{\mu\nu}(x)k^\mu k^\nu.
\end{align}
With the singular point removed, 
$\Gamma^+_0$ is a smooth differential manifold. The fiber space at $x\in \mathcal{M}$ is identified to be 
the momentum space for the photon,
\begin{align}
(\Gamma^+_0)_x:=\{k\in T_x\mathcal M\mid \mathcal{H}(x,k)=0,\ -Z_\mu k^\mu>0\},
\end{align}
and clearly one has
\begin{align}
\Gamma^+_0=\bigcup_{x\in\mathcal{M}}(\Gamma^+_0)_x.
\end{align}
It is important to note, however, that $\Gamma^+_0$ is still not the space of microstates, 
because it includes an extra temporal dimension from the base manifold $\mathcal{M}$. 

To establish a coordinate-independent notion of time, consider a set of observers 
whose proper velocities form a vector field $Z^\mu$ satisfying $Z \wedge \D Z = 0$, 
where $Z = g_{\mu\nu}Z^\mu \D x^\nu$.\footnote{For a detailed discussion of 
observer-dependent spacetime decomposition, see \cite{cai2023relativistic}.} 
By Frobenius theorem, this condition ensures the existence of a scalar field $t(x)$ such that each 
hypersurface $\mathcal{S}_t = \{x \in \mathcal{M} \mid t(x) = t\}$ is orthogonal to $Z^\mu$. 
These hypersurfaces serve as configuration spaces of simultaneous events. 
The space of microstates at time $t$ is then defined as
\begin{align}
\Sigma_t := \{(x,k) \in \Gamma^+_0 \mid t(x) = t\}.
\end{align}
This can also be viewed as a fiber bundle of which the fiber space 
(\ie the momentum space)
$(\Gamma^+_0)_x := \{k \in T_x\mathcal{M} \mid \mathcal{H}(x,k) = 0, -Z_\mu k^\mu > 0\}$ 
is attached at each point on the base manifold (the configuration space) $\mathcal{S}_t$:
\begin{align}
\Sigma_t = \bigcup_{x \in \mathcal{S}_t} (\Gamma^+_0)_x,\qquad {\text{dim}(\Sigma_t)=2d}.
\end{align}
It should be emphasized that we do not use $t$ to parametrize photon trajectories. As discussed above, the trajectory parameter carries physical information and therefore cannot be changed arbitrarily. Moreover, retaining the affine parameter $\lambda$ allows us to formulate photon dynamics independent of the choice of observer.

The tangent vector of the phase trajectory $X_\lambda=(x^\mu(\lambda),k^\mu(\lambda))$ of a photon is obtained directly 
from the geodesic equation,
\begin{align}
\dot{X}_\lambda=\frac{\D x^\mu}{\D\lambda}\frac{\partial}{\partial x^\mu}
+\frac{\D k^\mu}{\D\lambda}\frac{\partial}{\partial k^\mu}
&= k^\mu \frac{\partial}{\partial x^\mu}
- \varGamma^\mu{}_{\rho\sigma} k^\rho k^\sigma \frac{\partial}{\partial k^\mu} \\
&= k^\mu \left( \frac{\partial}{\partial x^\mu}
- \varGamma^\nu{}_{\alpha\mu} k^\alpha \frac{\partial}{\partial k^\nu} \right)
= k^\mu e_\mu =: \mathcal{L},
\end{align}
where $e_\mu := \partial_{x^\mu} - \varGamma^\nu{}_{\alpha\mu} k^\alpha \partial_{k^\nu}$ 
is the horizontal lift \cite{sarbach2014geo} of $\partial_{x^\mu}$. The Liouville vector field 
$\mathcal{L}$ is the Hamiltonian vector field associated with $\mathcal H$. To verify this, introduce the symplectic form
on the tangent bundle, which can be constructed by the exterior derivative 
of the Poincare 1-form $\Theta:=g_{\mu\nu}k^\nu\D x^\mu$,
\begin{align}
\Omega:=\D\Theta=g_{\mu\nu}\theta^\mu\wedge\D x^\nu,
\quad
\theta^\mu:=\D k^\mu +\varGamma\ind{\mu}{\rho\nu}k^\rho\D x^\nu.
\end{align}
The horizontal and vertical bases satisfy
\begin{align}
\label{eq:basis-relation1}
\D x^\mu(e_\nu)&=\delta\ind{\mu}{\nu},\quad 
\D x^\mu\left(\frac{\partial}{\partial k^\nu}\right)=0,\\
\label{eq:basis-relation2}
\theta^\mu(e_\nu)&=0,\quad\quad\  
\theta^\mu\left(\frac{\partial}{\partial k^\nu}\right)=\delta\ind{\mu}{\nu}.
\end{align}
Hence $\{\D x^\mu, \theta^\mu\}$ is the dual basis to $\{e_\mu, \partial_{k^\mu}\}$. A direct computation gives
\begin{align}
-\iota_{\mathcal{L}}\Omega = \Omega(\ \cdot \ , \mathcal{L}) = k_\mu \theta^\mu = \D \mathcal H,
\end{align}
where $\iota$ denotes the interior derivative, the above relation confirming that $\mathcal L$ is 
Hamiltonian vector field associated with $\mathcal{H}$. 

It is well known that the Hamiltonian 
vector field preserves the symplectic form, ensuring that the Lie derivative of 
$\Omega$ with respect to the Hamiltonian vector field $\mathcal{L}$ must vanish. 
This can be verified by use of the Cartan formula,
\begin{align}
\pounds_{\mathcal{L}}\Omega=(\D\iota_{\mathcal{L}}+\iota_{\mathcal{L}}\D)\Omega
=-\D\D\mathcal{H}+\iota_{\mathcal{L}}\D\D\Theta=0,
\end{align}
and the Hamiltonian function $\mathcal{H}$ is conserved on the integral 
curve of $\mathcal{L}$,
\begin{align}
\mathcal{L}(\mathcal{H})=\D\mathcal{H}(\mathcal{L})=\Omega(\mathcal{L},\mathcal{L})=0.
\end{align}
Since $\mathcal{H}$ is constant on $\Gamma_0^+$, it follows that $\mathcal{L}$ is 
tangent to $\Gamma_0^+$.
Let $\mathscr{F}^n(M)$ denote the space of $n$ forms on a manifold $M$. Then
$\Omega\in\mathscr{F}^2(T\mathcal{M})$. Let
$i:\Gamma_0^+\hookrightarrow T\mathcal{M}$ denote the inclusion map. It follows
that $i^*\Omega\in\mathscr{F}^2(\Gamma_0^+)$. Strictly speaking, $\Omega$ and
its pullback $i^*\Omega$ are distinct differential forms, since they are
defined on different manifolds. For notational simplicity, however, we do not
distinguish between them explicitly in the formulas below. Similarly, since $\mathcal{L}$ is tangent to $\Gamma_0^+$, we likewise regard it as a vector field on $\Gamma_0^+$.

The natural metric on the tangent bundle $T\mathcal{M}$ is the Sasaki metric 
\cite{sasaki1958differential}, 
\begin{align}\label{eq:def-sasaki}
\hat g=g_{\mu\nu}\D x^\mu\otimes\D x^\nu+g_{\mu\nu}\theta^\mu\otimes\theta^\nu.
\end{align}
Since $\hat g$ is block diagonal under the dual basis $\{\D x^\mu,\theta^\mu\}$, 
the determinant of $\hat g$ is the product of the determinants of the two blocks. 
Therefore, the volume element associated with $\hat g$ reads
\begin{align}
\vol_{T\mathcal M}&=g~ \D x^0\wedge\cdots \wedge\D x^{d}\wedge \theta^0
\wedge\cdots \wedge\theta^{d}\notag\\
&=g~ \D x^0\wedge\cdots \wedge\D x^{d}\wedge \D k^0\wedge\cdots \wedge\D k^{d},
\end{align}
where $g=-\det g_{\mu\nu}$. 

The determinant of the spacetime metric can be written as
\begin{align}
-g=\frac{1}{(1+d)!}\left(\prod_{i=0}^{d} g_{\nu_i\mu_i} \right)\varepsilon^{\nu_0\nu_1\cdots \nu_{d}} 
\varepsilon^{\mu_0\mu_1\cdots \mu_{d}}.
\end{align}
Therefore,\begin{align}
\vol_{T\mathcal{M}}&=-\frac{1}{(1+d)!}\left(\prod_{i=0}^{d} g_{\nu_i\mu_i} \right)
\varepsilon^{\nu_0\nu_1\cdots \nu_{d}} \varepsilon^{\mu_0\mu_1\cdots \mu_{d}}\D x^0\wedge\cdots 
\wedge\D x^{d}\wedge \theta^0\wedge\cdots \wedge\theta^{d}\notag\\
&=-\frac{1}{(d+1)!}\left(\prod_{i=0}^{d} g_{\nu_i\mu_i}\right)\D x^{\mu_0}\wedge\cdots \wedge\D x^{\mu_d}\wedge 
\theta^{\nu_0}\wedge\cdots \wedge\theta^{\nu_d}\notag\\
&=-\frac{(-1)^{(d+1)(d+2)/2}}{(d+1)!}\left(\prod_{i=0}^{d} g_{\nu_i\mu_i}\right)
\theta^{\nu_0}\wedge\D x^{\mu_0}\wedge\cdots \wedge\theta^{\nu_d}\wedge \D x^{\mu_d}\notag\\
&=-\frac{(-1)^{(d+1)(d+2)/2}}{(d+1)!}\Omega\wedge\cdots \wedge\Omega.
\end{align}
Since $\mathcal{L}$ preserves the symplectic form, the above equation implies that
it is also volume-preserving, \ie
\begin{align}\label{eq:Liouville-thm-tangent}
    \pounds_\mathcal{L}\vol_{T\mathcal{M}}=0.
\end{align}
This result gives the Liouville theorem on the tangent bundle.

To examine the causal structure of the light cone bundle, consider the vector field
\begin{align}
\hat n := k^\mu \frac{\partial}{\partial k^\mu},
\end{align}
which points along the radial direction in momentum space, i.e. the direction 
of the photon momentum itself. Using $\hat g(\hat n,\cdot)=k_\mu\theta^\mu=\D \mathcal H$ 
and the fact that $\mathcal H$ is constant on $\Gamma_0^+$, we have 
$\hat g(\hat n,X)=0$ for every vector $X$ tangent to $\Gamma_0^+$. Thus $\hat n$ 
is normal to $\Gamma_0^+$. Meanwhile, $\hat g(\hat n,\hat n)=g_{\mu\nu}k^\mu k^\nu=0$, 
so $\hat n$ is null and hence also tangent to $\Gamma_0^+$. Consequently, the 
induced metric $\hat h := i^*\hat g$ on $\Gamma_0^+$ is degenerate, and its 
associated volume element vanishes. This degeneracy precludes the definition of a scalar distribution function on a manifold and constitutes the central geometric obstacle that we must overcome.

To proceed, we recall a general fact from kinetic theory. On any manifold, a distribution is described by a full-rank form $\eta$, whose evolution is governed by the standard transport equation (see Appendix~\ref{app:distribution-on-manifold}). If there exists a volume element $\vol_M$ that is invariant under the Hamiltonian flow, then it serves as a stationary background, and the distribution can be decomposed as $\eta = f \vol_M$, with all dynamics encoded in the scalar function $f$. Thus, for our purposes, it suffices to construct a nonzero volume element on $\Gamma_0^+$, even one not compatible with $\hat h$, so that the degeneracy of $\hat h$ no longer prevents the definition of a scalar distribution function. In this section, we construct such a volume element, using a method analogous to the Hodge dual, and prove that it satisfies Liouville's theorem.

{Such a volume element cannot be obtained by the standard interior product method. Since the normal vector $\hat n$ is also tangent to the light cone bundle, it follows from Lem.~\ref{lem:restriction} that the induced volume element
\begin{align}
    \iota_{\hat n}\vol_{T\mathcal{M}}
\end{align}
vanishes on the light cone bundle. Nevertheless, despite the fact that $\hat n$ is both normal and tangent to $\Gamma_0^+$, one may proceed in a manner analogous to the Hodge dual of the normal covector $\D\mathcal{H}$ to define a nonvanishing volume element on the light cone bundle, namely, as the solution of}
\begin{align}\label{eq:relation-volumes}
    \vol_{T\mathcal{M}}=(-1)^{d+1}\D\mathcal{H}\wedge \vol_{\Gamma_0^+}.
\end{align}
The solution $\vol_{\Gamma_0^+}$ on $T\mathcal{M}$ is not unique, since one may add to it any term containing $\D\mathcal H$ as a factor. However, such terms vanish upon restriction to $\Gamma_0^+$, where $\D\mathcal H=0$ on tangent vectors. Thus the restriction of $\vol_{\Gamma_0^+}$ to $\Gamma_0^+$ is unique. The explicit solution is
\begin{align}\label{eq:volume-element-lightcone}
    \vol_{\Gamma_0^+} := &\frac{g}{k_0}\D x^0 \wedge \ldots \wedge \D x^{d} \wedge \D k^1 \wedge \ldots \wedge \D k^{d}\notag \\
    =& \frac{g}{k_0}\D x^0 \wedge \ldots \wedge \D x^{d} \wedge \theta^1 \wedge \ldots \wedge \theta^{d}.
\end{align}
It is well known that the invariant volume element on spacetime is given by
\begin{align}
    \vol_{\mathcal{M}}=\sqrt{g}\,\D x^0\wedge\ldots\wedge\D x^{d}.
\end{align}
Let $\pi:\Gamma^+_0\rightarrow\mathcal{M}$ denote the canonical projection, defined by $\pi(x,k)=x$. Consequently, the volume element on the light cone bundle decomposes as
\begin{align}
    \vol_{\Gamma_0^+}=\pi^*(\vol_{\mathcal{M}})\wedge\varpi,
\end{align}
where $\varpi:=(\sqrt{g}/k_0 )\D k^1\wedge\ldots \wedge\D k^d$ is the volume element on the momentum space. Strictly speaking, since the spacetime manifold $\mathcal{M}$ and the light cone bundle are distinct manifolds, the form $\vol_{\mathcal{M}}$ must be pulled back by $\pi$ before it can be wedged with $\varpi$. For notational simplicity, however, we suppress this distinction below and write
\begin{align}\label{eq:decompose-volume-gamma}
    \vol_{\Gamma_0^+}=\vol_{\mathcal{M}}\wedge\varpi.
\end{align}

\begin{theorem}
\label{thm:Liouville-thm-lightcone}
\textup{\textbf{Liouville's theorem on the light cone bundle}: } 
The Liouville vector field $\mathcal{L}$ preserves the volume element 
$\vol_{\Gamma_0^+}$ on the light cone bundle, i.e.
\begin{align}
\pounds_\mathcal{L}\vol_{\Gamma_0^+}=0.	
\label{Lvlcone}
\end{align}
\begin{proof}
Combining the Liouville's theorem \eqref{eq:Liouville-thm-tangent} on the tangent bundle 
and the relation \eqref{eq:relation-volumes}, we have
\begin{align}\label{eq:thm-1}
0=(\pounds_\mathcal{L}\D\mathcal{H})\wedge\vol_{\Gamma_0^+}
+\D\mathcal{H}\wedge\pounds_\mathcal{L}\vol_{\Gamma_0^+}.
\end{align}
The Lie derivative of $\D\mathcal{H}$ along $\mathcal{L}$ is zero, as can be 
shown by use of the Cartan formula
\begin{align}
\pounds_\mathcal{L}\D\mathcal{H}=(\D\iota_\mathcal{L}
+\iota_\mathcal{L}\D)\D\mathcal{H}=\D(\mathcal{L}(\mathcal{H}))=0.
\end{align}
Therefore, eq.~\eqref{eq:thm-1} can be rewritten as 
\begin{align}
0=\D\mathcal{H}\wedge\pounds_\mathcal{L}\vol_{\Gamma_0^+}.
\end{align}
Let $\{\mathcal{X}_1,\mathcal{X}_2,\cdots ,\mathcal{X}_{2d+1}\}$ be an 
arbitrary set of vectors tangent to $\Gamma_0^+$, and $l$ be a vector 
which is not tangent to $\Gamma_0^+$. Since $\D\mathcal{H}$ is a normal covector 
of $\Gamma_0^+$, it implies
\begin{align}\label{eq:thm-2}
0=\D\mathcal{H}\wedge\pounds_\mathcal{L}\vol_{\Gamma_0^+}(l,\mathcal{X}_1,
\mathcal{X}_2,\cdots ,\mathcal{X}_{2d+1})
=\D\mathcal{H}(l)\cdot\pounds_\mathcal{L}\vol_{\Gamma_0^+}
(\mathcal{X}_1,\mathcal{X}_2,\cdots ,\mathcal{X}_{2d+1}),
\end{align}
and since $l$ is transverse to $\Gamma_0^+$, we have
\begin{align}\label{thm-3}
\D\mathcal{H}(l)\neq 0.
\end{align}
Combining eqs.~\eqref{eq:thm-2} and \eqref{thm-3}, eq.~\eqref{Lvlcone} follows.
\end{proof}
\end{theorem}

Liouville's theorem allows the volume element $\vol_{\Gamma_0^+}$ to be 
considered as a stationary background, ensuring that the evolution of 
$\eta = f \vol_{\Gamma_0^+}$ is entirely due to the evolution of the 
scalar part $f$. Therefore, Liouville's theorem is crucial for the 
scalarization of the evolution equation of $\eta$. 

However, this volume element alone is not sufficient for relativistic kinetic theory. Physical observables are not given directly by the scalar distribution function $f$; rather, they arise as fluxes through spacelike hypersurfaces. Such fluxes are naturally described by the inner product between the density current and the hypersurface's unit normal, which relies on Gauss's theorem and therefore, requires compatibility between the volume element and the induced metric. Since $\vol_{\Gamma_0^+}$ is not compatible with $\hat h$, the standard flux description is not directly available. In the next section, we will first use the scalarization approach to derive the tensorial Boltzmann equation and then develop an alternative route to connect the flux to the current vector.

\section{Boltzmann equation for photons and its scalarization}
\label{sec:boltzmann}

The evolution operator for photons on the light cone bundle $\Gamma_0^+$ 
is generated by the Liouville vector field
\begin{align}
    U_\lambda := \re^{\mathcal{L} \lambda}.
\end{align}
The phase trajectories $X_\lambda$ of a photon are governed by the 
evolution operator, \ie $X_\lambda = U_\lambda(X_0)$, where $X_0:=X_\lambda|_{\lambda=0}$.
In the presence of interactions, the interaction part of the Hamiltonian
can create and annihilate photons. We therefore introduce the full-rank form $\mathscr{C}_\lambda\in\mathscr{F}^{2d+1}(\Gamma^+_0)$ to describe photon creation and annihilation for any $\lambda\in\mathbb{R}$.
Along a photon trajectory, $\mathscr{C}_\lambda$ represents the net photon-number source density per unit
affine parameter $\lambda$. From an observer's perspective, the value of this internal parameter at which a photon crosses $\Sigma_t$ is not directly resolved, and one is therefore primarily interested in the integral over $\lambda$:
\begin{align}
    \mathscr{C}:=\int_{\mathbb{R}}\D\lambda\,\mathscr{C}_\lambda.
\end{align}
It is typically represented by scattering integrals. Physically, we assume that the total creation and annihilation contributions throughout spacetime are finite, so that this integral converges absolutely.

The distribution of these photons when their affine parameter is $\lambda$ can be 
described by a full-rank form $\eta_\lambda{\in\mathscr{F}^{2d+1}(\Gamma^+_0)}$. The evolution equation 
of $\eta_\lambda$ is given by
\begin{align}\label{eq:tensor-Boltzmann-lambda}
    \partial_\lambda \eta_\lambda + \pounds_\mathcal{L} \eta_\lambda = \mathscr{C}_\lambda,
\end{align}
as derived in Appendix~\ref{app:distribution-on-manifold}. It should be emphasized that $\eta_\lambda$ is not itself the distribution measured by an observer, since it is defined on the entire light cone bundle $\Gamma_0^+$, which still contains a time dimension. From an observer's perspective, the physical distribution should instead be defined on a hypersurface $\Sigma_t$ at fixed time $t$. It is given by the flux of $\eta_\lambda$ through $\Sigma_t$.
According to Thm.~\ref{thm:flux}, the flux of $\eta_\lambda$ reads
\begin{align}\label{eq:distribution-per-lambda}
    \xi_\lambda = \iota_\mathcal L \eta_\lambda,
\end{align}
which represents the distribution of the intersection points between $\Sigma_t$ and the corresponding phase trajectories per unit affine parameter $\lambda$. However, an observer detects only the intersection points on $\Sigma_t$ and is insensitive to the value of the internal parameter at which each photon crosses $\Sigma_t$.
Therefore, the physical distribution on $\Sigma_t$ is obtained by integrating eq.~\eqref{eq:distribution-per-lambda} over the affine parameter $\lambda$,
\begin{align}\label{eq:def-flux-photons}
    \xi = \int_{\mathbb{R}} \D \lambda \, \xi_\lambda = \iota_\mathcal{L} \eta,
\end{align}
where $\eta$ is defined by
\begin{align}
    \eta := \int_{\mathbb{R}} \D \lambda \, \eta_\lambda.
\end{align}
If the causal structure of spacetime is well behaved, in the sense that no null geodesic can intersect the same spacelike hypersurface more than once, then Thm.~\ref{thm:flux-convergence}, together with our integrability assumptions on $\eta_\lambda$ and $\mathscr{C}_\lambda$, implies that the integral of $\xi_\lambda$ over $\lambda$ converges absolutely.
Although $\eta$ is still not a physical distribution, it is related to the physical state density distribution $\xi$ through \eqref{eq:def-flux-photons}.
Since $\eta$ is a full-rank form on $\Gamma_0^+$ equipped with a volume element satisfying Liouville's theorem, it is better suited for scalarization. We can derive the evolution equation for $\eta$ firstly, then use eq.~\eqref{eq:def-flux-photons} to extend this result to $\xi$. 

Because photons are either created or annihilated during a finite temporal interval, or from past infinity to future infinity, for a finite region of spacetime one naturally has
\begin{align}\label{eq:boundary-eta}
    \eta_{+\infty}|_{(x,k)} = \eta_{-\infty}|_{(x,k)} = 0,\qquad \text{for }\forall (x,k)\in\Gamma^+_0.
\end{align}
The tensorial Boltzmann equation for photons can be derived from integrating eq.~\eqref{eq:tensor-Boltzmann-lambda} and applying the boundary conditions in eq.~\eqref{eq:boundary-eta},
\begin{align}\label{eq:tensor-Boltzmann}
    \pounds_\mathcal{L} \eta = \mathscr{C}.
\end{align}
The similar approach dealing with $\eta_\lambda$ has been used in relativistic stochastic mechanics \cite{cai2023relativistic2}. 

With the volume element $\vol_{\Gamma_0^+}$ constructed in the previous section, which satisfies Liouville's theorem, the scalarization of \eqref{eq:tensor-Boltzmann} becomes possible. Writing the scalar distribution function $f$ and the source $\mathcal{C}$ as
\begin{align}
    \eta=f\vol_{\Gamma_0^+},\qquad \mathscr{C}=\mathcal{C}\vol_{\Gamma_0^+}.
\end{align}
and using $\pounds_\mathcal L \vol_{\Gamma_0^+}=0$, we obtain
\begin{align}
    \pounds_\mathcal{L}\eta=(\pounds_\mathcal{L}f)\vol_{\Gamma_0^+}+f\pounds_\mathcal{L}\vol_{\Gamma_0^+}=\mathcal{L}(f)\vol_{\Gamma_0^+}=\mathcal{C}\vol_{\Gamma_0^+}.
\end{align}
Thus the tensorial Boltzmann equation reduces to the scalar form,
\begin{align}\label{eq:scalar-Boltzmann}
    \mathcal{L}(f)=\mathcal{C}.
\end{align}
This is precisely the familiar Boltzmann equation, and it has the same form as in the massive case.

The tensorial Boltzmann equation \eqref{eq:tensor-Boltzmann} can be expressed in the form of a conservation equation. According to the Cartan formula, the Lie derivative of the distribution $\eta$ with respect to the vector field $\mathcal{L}$ can be written as the exterior derivative of $\xi$,
\begin{align}\label{eq:lie-derivative-eta}
    \pounds_\mathcal{L}\eta = \D \iota_\mathcal{L}\eta = \D \xi.
\end{align}
Substituting eq.~\eqref{eq:lie-derivative-eta} into eq.~\eqref{eq:tensor-Boltzmann}, we obtain,
\begin{align}\label{eq:tensor-conservation-eq}
    \D \xi = \mathscr{C}.
\end{align}
To consider the difference in total particle numbers between times $t_1$ and $t_2$, Stokes' theorem allows us to convert this difference into an integral over the region $\Gamma[t_1,t_2]$ between the surfaces $\Sigma_{t_1}$ and $\Sigma_{t_2}$:
\begin{align}\label{eq:difference-particle-number1}
    N_{t_2} - N_{t_1} = \int_{\Sigma_{t_2}} \xi - \int_{\Sigma_{t_1}} \xi = \int_{\Gamma[t_1,t_2]} \D \xi = \int_{\Gamma[t_1,t_2]} \mathscr{C}.
\end{align}
Thus, $\D \xi$ represents the change in particle number per unit phase volume and unit time, and eq.~\eqref{eq:tensor-conservation-eq} indicates that the change in particle number is equal to the particle number generated by the source $\mathscr{C}$. This is precisely a conservation law with source.

Typically, in relativity, density quantities are defined as the inner product between a density current and the observer's velocity vector, and are therefore observer dependent. The distribution $\xi$ describes the particle density of relativistic particles. This suggests that $\xi$ should also be observer dependent. The configuration space $\mathcal{S}_t$ can be regarded as the hypersurface where $t(x)$ is constant. Since the spacetime metric $g_{\mu\nu}$ is nondegenerate, the normal covector of $\mathcal{S}_t$ must be proportional to $\D t$, \ie
\begin{align}
    Z_\mu \propto \partial_\mu t.
\end{align}
The space of microstates $\Sigma_t$ can also be regarded as the hypersurface where $\hat{t}(x,k):=t(x)$ is constant. This implies that $Z_\mu \D x^\mu$ is proportional to $d\hat t=\partial_\mu t\,\D x^\mu$ and hence, is a normal covector of $\Sigma_t$. Let
\begin{align}
    \mathcal{Z}=Z^\mu e_\mu,
\end{align}
Using the defining properties of $e_\mu$, one finds $\hat h(\ \cdot\ ,\mathcal{Z})=Z_\mu \D x^\mu$. Therefore, $\mathcal{Z}$ is a normal vector of the space of microstates $\Sigma_t$ associated with the observer $Z^\mu$ introduced in Sec.~\ref{sec:light-cone-bundle}.

Since $\hat h$ is degenerate, the unit normal vector of $\Sigma_t$ is not unique. For any scalar field $\alpha$,
\begin{align}
    \mathcal{Z}+\alpha\hat n
\end{align}
is still the unit normal vector of $\Sigma_t$. However, since $\hat n$ is normal to $\Gamma_0^+$ and $\mathcal{L}$ is tangent to $\Gamma_0^+$, we have $\hat h(\hat n,\mathcal{L})=0$. Hence the $\alpha$ term does not contribute to $\hat h(\mathcal{Z},\mathcal{L})$, and we may set $\alpha=0$ without loss of generality. The Liouville vector field $\mathcal{L}$ can be decomposed as 
\begin{align}
    \mathcal{L}=-\hat h(\mathcal{Z},\mathcal{L})\mathcal{Z}+l.
\end{align}
Using the decomposition of $\mathcal L$ into normal and tangential parts relative to $\Sigma_t$, Lem.~\ref{lem:restriction} implies that $\iota_l\eta$ restricted on $\Sigma_t$ is zero. Hence, from $\xi=\iota_\mathcal L\eta$,
\begin{align}
    \xi = -\hat h(\mathcal Z,\mathcal L)\,\iota_\mathcal Z\eta \qquad \text{on } \Sigma_t.
\end{align}
With $\eta = f\vol_{\Gamma_0^+}$ and $\mathcal J := f\mathcal L$, this becomes
\begin{align}\label{eq:xi-Z}
    \xi = -\hat h(\mathcal Z,\mathcal L)f\,\iota_\mathcal Z\vol_{\Gamma_0^+}
        = -\hat h(\mathcal Z,\mathcal J)\,\vol_{\Sigma_t},
\end{align}
where $\vol_{\Sigma_t}:=\iota_\mathcal Z\vol_{\Gamma_0^+}$ is the volume element of $\Sigma_t$.
Although the induced metric $\hat h$ and the volume element $\vol_{\Gamma_0^+}$ on the light cone bundle are not directly compatible, eq.~\eqref{eq:xi-Z} shows that the particle density can nevertheless be described by the metric pairing $\hat h(\mathcal Z,\mathcal J)$ multiplied by the volume element $\vol_{\Sigma_t}=\iota_\mathcal Z\vol_{\Gamma_0^+}$. Thus, $\hat h$ and $\vol_{\Gamma_0^+}$ together provide the geometric structure of the light cone bundle. We denote this geometric structure by $(\Gamma_0^+,\vol_{\Gamma_0^+},\hat h)$. 

The relation between the induced metric and the volume element revealed by eq.~\eqref{eq:xi-Z} renders the expressions for physical observables formally identical to those for massive particles. To make this explicit, we first compute the components of $\xi$ and $\D\xi$ in detail, and then use these relations to derive the conservation law. Indeed, using the Sasaki metric,
\begin{align}
    \hat{h}(\mathcal{Z},\mathcal{J})=Z_\mu k^\mu f,
\end{align}
so that, the state density distribution $\xi$ can also be connected with the scalar distribution function $f$
\begin{align}\label{eq:xi-f}
    \xi=-Z_\mu k^\mu f\vol_{\Sigma_t}.
\end{align}
Despite the fact that there is no covariant derivative on $(\Gamma_0^+,\vol_{\Gamma_0^+},\hat h)$ since $\hat h$ is degenerate, the divergence of a vector field $\mathcal{A}$ on $\Gamma_0^+$ still can be defined by the volume element $\vol_{\Gamma_0^+}$ (see the definition 9.8 in \cite{lee2022manifolds}):
\begin{align}
    \mathrm{div}\mathcal{A}\ \vol_{\Gamma_0^+}=\pounds_{\mathcal{A}}\vol_{\Gamma_0^+}.
\end{align}
Using the scalarization $\eta = f\vol_{\Gamma_0^+}$, the definition $\mathcal J := f\mathcal L$, and Cartan's formula together with $\D\vol_{\Gamma_0^+}=0$ (since $\vol_{\Gamma_0^+}$ is a top-rank form), we obtain
\begin{align}\label{eq:divergence-J-dxi}
    \D \xi = \D\iota_{\mathcal L}\eta = \D (\iota_{\mathcal J}\vol_{\Gamma_0^+}) = \pounds_{\mathcal J}\vol_{\Gamma_0^+} = \mathrm{div}\,\mathcal J\; \vol_{\Gamma_0^+}.
\end{align}
Together with the conservation equation $\D\xi = \mathscr C$, this yields the current conservation form
\begin{align}
    \mathrm{div}\,\mathcal J = \mathcal C.
\end{align}
Integrating this equation over $\Gamma[t_1,t_2]$ and using the flux expression $\xi = -\hat h(\mathcal Z,\mathcal J)\vol_{\Sigma_t}$ recovers the integral form
\begin{align}
    N_{t_2} - N_{t_1} &= -\int_{\Sigma_{t_2}}\hat{h}(\mathcal{Z},\mathcal{J})\vol_{\Sigma_{t_2}} + \int_{\Sigma_{t_1}}\hat{h}(\mathcal{Z},\mathcal{J})\vol_{\Sigma_{t_1}} \notag\\
    &= \int_{\Gamma[t_1,t_2]} \mathrm{div}\,\mathcal J\; \vol_{\Gamma_0^+} = \int_{\Gamma[t_1,t_2]} \mathcal C\,\vol_{\Gamma_0^+}.
\end{align}
In the next section, we apply the same strategy to define the corresponding macroscopic currents and derive their conservation laws.

\section{Currents on spacetime manifold}
\label{sec:current}

A central application of kinetic theory is to provide a statistical foundation for hydrodynamics. The fiber bundle structure permits integration only over the fiber component of a full-rank differential form, thereby inducing a density on the base manifold from which hydrodynamic quantities can be constructed.

An observable quantity $\mathcal{O}$ of photons can be regarded as a scalar field on $\Gamma_0^+$, its expectation in time $t$ is given by the integral of the product of $\mathcal{O}$ and $\xi$ over the space of microstates $\Sigma_t$,
\begin{align}\label{eq:expectation1}
    \bar{\mathcal{O}}_t=\int_{\Sigma_t}\mathcal{O}\xi.
\end{align}
As in eq.~\eqref{eq:decompose-volume-gamma}, the volume element of the space of microstates factorizes as the wedge product of the volume elements of the configuration space $\mathcal{S}_t$ and the momentum space $(\Gamma_0^+)_x$,
\begin{align}\label{eq:decompose-volume-sigma}
    \vol_{\Sigma_t}=\iota_{\mathcal{Z}}\vol_{\Gamma_0^+}=(\iota_Z \vol_{\mathcal M})\wedge\varpi=\vol_{\mathcal{S}_t}\wedge\varpi.
\end{align}
The second equality in the above equation follows from Lem.~\ref{lem:decomposition}.
Utilizing eqs.~\eqref{eq:xi-f} and \eqref{eq:decompose-volume-sigma}, the expectation \eqref{eq:expectation1} can be rewritten as the integral of the inner product between the current for $\mathcal{O}$ and the observer $Z^\mu$ over the configuration space
\begin{align}\label{eq:expectation2}
    \bar{\mathcal{O}}_t=-\int \vol_{\Sigma_t} \hat{h}(\mathcal Z,\mathcal J)\mathcal{O}  =-\int_{\mathcal{S}_t}\vol_{\mathcal{S}_t}Z_\mu\int_{(\Gamma_0^+)_x}\varpi k^\mu \mathcal{O} f=-\int_{\mathcal{S}_t}\vol_{\mathcal{S}_t}Z_\mu\mathcal{O}^\mu,
\end{align} 
where the current is defined by
\begin{align}\label{eq:current}
    \mathcal{O}^\mu=\int_{(\Gamma_0^+)_x}\varpi f k^\mu \mathcal{O}.
\end{align}
In the massive-particle case, $\bar{\mathcal{O}}_t$ is typically defined by the first equality in eq.~\eqref{eq:expectation2}, thereby ensuring that the definition of the macroscopic current for massless particles is formally consistent with that for massive particles. Let $V$ be an arbitrary region within the spacetime manifold $\mathcal{M}$, and let $\Gamma=\pi^{-1}[V]$ be its preimage on the light cone bundle $\Gamma_0^+$, where $\pi:\Gamma_0^+\rightarrow\mathcal{M}$ is the projection mapping. Gauss's theorem and Stokes's theorem imply 
\begin{align}
    \int_V\vol_V \nabla_\mu\mathcal{O}^\mu& =\int_{\partial V}\vol_{\partial V}v_\mu\mathcal{O}^\mu=-\int_{\partial\Gamma}\vol_{\partial\Gamma}v_\mu k^\mu\mathcal{O}f\xlongequal[]{\text{Eq.~}\eqref{eq:xi-f}}\int_{\partial \Gamma}\mathcal{O}\xi =\int_{\Gamma}\D(\mathcal{O}\xi) \notag\\
    &=\int_\Gamma \D(\iota_{\mathcal{L}}\mathcal{O}\eta)=\int_\Gamma\pounds_\mathcal{L}(\mathcal{O}f \vol_{\Gamma_0^+})=\int_{V}\vol_{V}\int_{(\Gamma_0^+)_x}\varpi\mathcal{L}(f\mathcal{O}),
\end{align}
where $v^\mu$ is unit normal to $\partial V$. 
In the second equality above, we have used the following decomposition of the boundary volume element, analogous to eq.~\eqref{eq:decompose-volume-sigma}:
\begin{align}
    \vol_{\partial \Gamma}=\iota_{v^\mu e_\mu} \vol_{\Gamma^+_0}=(\iota_v \vol_\mathcal{M})\wedge\varpi=\vol_{\partial V}\wedge \varpi.
\end{align}
Since $V$ is arbitrary, the divergence of $\mathcal{O}^\mu$ follows
\begin{align}\label{eq:divergence}
    \nabla_\mu\mathcal{O}^\mu=\int\varpi\mathcal{L}(f\mathcal{O}).
\end{align}
For particle number, set $\mathcal O=1$ in eq.~\eqref{eq:current}. The particle current is then
\begin{align}
    N^\mu := \int \varpi f k^\mu,
\end{align}
and its divergence follows from eq.~\eqref{eq:divergence},
\begin{align}\label{eq:divergence-particle-current} 
    \nabla_\mu N^\mu = \int \varpi \mathcal L(f).
\end{align}
Similarly, the photon energy $\omega = -Z_\mu k^\mu$ defines an energy current,
\begin{align}
    \omega^\mu := -Z_\nu \int \varpi f k^\nu k^\mu = -Z_\nu T^{\mu\nu},
\end{align}
where
\begin{align}
    T^{\mu\nu} := \int \varpi f k^\nu k^\mu
\end{align}
is the energy-momentum tensor. Applying eq.~\eqref{eq:divergence} to $\omega^\mu$ and using the identity
$\mathcal L(Z_\mu k^\mu)=k^\mu k^\nu\nabla_\mu Z_\nu$ gives
\begin{align}\label{eq:divergence-energy-current1} 
    \nabla_\mu \omega^\mu&=-\int\varpi\mathcal{L}(f Z_\mu k^\mu)=-Z_\mu\int\varpi k^\mu\mathcal{L}(f)-\int\varpi f\mathcal{L}(Z_\mu k^\mu)\notag\\ &=-Z_\mu\int\varpi k^\mu\mathcal{L}(f)-\nabla_\mu Z_\nu\int\varpi f k^\mu k^\nu=-Z_\mu\int\varpi k^\mu\mathcal{L}(f)-T^{\mu\nu}\nabla_\mu Z_\nu. 
\end{align}
Since the energy current is the contraction between the energy-momentum tensor and the observer, the divergence of $\omega^\mu$ can also be expressed as
\begin{align}\label{eq:divergence-energy-current2} 
    \nabla_\mu \omega^\mu&=-Z_\nu\nabla_\mu T^{\mu\nu}-T^{\mu\nu}\nabla_\mu Z_\nu. 
\end{align}
Combining eqs.~\eqref{eq:divergence-energy-current1} and \eqref{eq:divergence-energy-current2}, we obtain the divergence of the energy-momentum tensor:
\begin{align}\label{eq:divergence-energy-current} 
    \nabla_\nu T^{\mu\nu}=\int\varpi k^\mu\mathcal{L}(f). 
\end{align}
For a photon gas, the single-particle entropy is
\begin{align} 
    S= -\log \left( \frac{h^df}{f^{\ast} \mathfrak{g}} \right) + \frac{\mathfrak{g} \log f^{\ast}}{\varsigma h^d f}, 
\end{align}
where $f^*:=1+\varsigma \mathfrak{g}^{-1}h^df$, $h$ is Planck constant, $\mathfrak{g}=d-1$ is the quantum degeneracy of photons, and $\varsigma=1$ represents Bose-Einstein statistics \cite{cercignani2002relativistic}. The entropy current then follows from eq.~\eqref{eq:current},
\begin{align} 
    S^\mu=-\int \varpi k^{\mu} f \left[ \log \left( \frac{h^d f }{f^{\ast} \mathfrak{g}} \right) - \frac{\mathfrak{g} \log f^{\ast}}{\varsigma h^d f} \right], 
\end{align}
and the divergence of the entropy current can also be derived by eq.~\eqref{eq:divergence},
\begin{align}\label{eq:divergence-entropy-current}
    \nabla_\mu S^\mu=-\int\varpi \mathcal{L}(f) \log\frac{h^df}{\mathfrak{g}f^*}.
\end{align}

If there are no other forms of matter except photons in the curved spacetime, the Boltzmann equation for photons \eqref{eq:scalar-Boltzmann} simplifies to the Liouville equation:
\begin{align} 
    \mathcal{L}(f)=0. 
\end{align}
As a result, the divergences of the particle current \eqref{eq:divergence-particle-current}, the energy-momentum tensor \eqref{eq:divergence-energy-current}, and the entropy current \eqref{eq:divergence-entropy-current} all vanish. This corresponds to the free-streaming limit, where photons propagate along geodesics without scattering or absorption.

When considering the interaction between photons and other forms of matter, the photon particle number becomes nonconservative, as photons are radiative particles. However, the total energy-momentum tensor of both matter and photons remains a conserved quantity. Additionally, the total entropy must monotonically increase, in accordance with the H theorem. The explicit verification of these statements requires the symmetry properties of the scattering integral, which are beyond the scope of the present geometric construction. These aspects will be addressed in future work.

\section{Conclusion and prospect}
\label{sec:conclusion}
In this paper, we have developed the geometric foundation for a relativistic kinetic theory of photons in curved spacetime. The central difficulty is that the induced metric on the future light cone bundle is degenerate, which renders the standard volume element vanishing and prevents the usual construction from carrying over directly from the massive case. To overcome this, we employed a construction analogous to the Hodge dual to define a nonzero volume element $\vol_{\Gamma_0^+}$ on the light cone bundle. Although this volume element is not compatible with the degenerate induced metric $\hat h$, it satisfies Liouville's theorem and thus serves as a valid stationary background for scalarizing the distribution as $\eta = f\vol_{\Gamma_0^+}$.

The volume element and the induced metric are nevertheless not independent. Eq.~\eqref{eq:xi-Z} shows that, once a spacelike hypersurface is specified, the particle density is determined by the inner product of the particle current with its normal vector, evaluated with the induced metric $\hat{h}$. The hypersurface volume element is obtained by contracting $\vol_{\Gamma_0^+}$ with the same normal vector. This relation between the volume element and the induced metric is precisely what makes the particle, energy, and entropy currents take the same form as in the massive case. Thus, $\hat{h}$ and $\vol_{\Gamma_0^+}$ jointly define the geometry of the light cone bundle.

Building on this geometric structure $(\Gamma_0^+, \vol_{\Gamma_0^+}, \hat h)$, we derived the fully covariant Boltzmann equation for photons in its tensorial form and then reduced it to the scalar equation $\mathcal L(f)=\mathcal C$. Using this kinetic equation, we obtained the macroscopic currents and computed their covariant divergences. Unsurprisingly, their expressions are identical to those for massive particles. The framework developed here may serve as a basis for future applications to strong-field photon transport, such as black hole shadow and accretion disk studies \cite{qi2023gravitational,zhao2024lensing,Cunha:2018acu,Gralla:2019xty,kato2008black}.
The present work has concentrated on the geometric aspects. A complete kinetic theory also requires a detailed treatment of scattering integrals, the derivation of the H theorem, and the explicit coupling to matter sources. These extensions will be addressed in forthcoming work.

\section*{Acknowledgment}
This work is supported by the National Natural Science Foundation of China (Grants No.~12275216, No.~12275138, No.~12247103).

{
\footnotesize\itemsep=0pt
\bibliographystyle{utphys}
\bibliography{reference.bib}
}

\begin{appendices}
    \section{Statistical equation on a manifold}
\label{app:distribution-on-manifold}

Let $M$ be an $n$-dimensional manifold which may be either Riemannian or symplectic. 
Suppose there are many particles 
moving within $M$. The distribution of these particles at time $t$ is described 
by a full-rank form $\eta_t$. The number of particles in a region $V$ of $M$ is given by the integral of $\eta_t$ over $V$,
\begin{align}
N_t = \int_V \eta_t.
\end{align}
Assume that these particles move along the integral curves of a vector field $L$. We denote the time evolution operator as
\begin{align}
U_t := \re^{Lt}.
\end{align}
Additionally, there are particle sources and sinks in $M$. This is described by a full-rank form $\mathcal{C}_t$. The integral of $\mathcal{C}_t$ over $V$ represents the number of particles created in $V$ minus the number of particles annihilated in $V$ per unit time. Let $V$ also move along the integral curves of $L$, \ie ,
\begin{align}
    V_t = U_t[V].
\end{align}
Then, the particle number in $V_t$ at time $t$ is the sum of the particle number in $V_0=V$ at time $0$ and the particles generated by the source $\mathcal{C}_t$ from $0$ to $t$,
\begin{align}\label{eq:particle-number-Vt}
    \int_{V_t} \eta_t = \int_{V} \eta_0+\int_0^t \D\tau \int_{V_\tau} \mathcal{C}_\tau.
\end{align}
Since the pullback induced by $U_t:V\rightarrow V_t$ maintains the integral of a full-rank form, \ie ,
\begin{align}
    \int_{V_t}\eta_t=\int_{V} U^*_t \eta_t,\qquad \int_{V_\tau}\mathcal{C}_{\tau}=\int_{V} U^*_\tau \mathcal{C}_{\tau},
\end{align}
and $V$ is arbitrary, eq.~\eqref{eq:particle-number-Vt} can be written as
\begin{align}\label{eq:particle-number-Vt-2}
    U^*_t\eta_t=\int^t_0\D\tau U^*_\tau\mathcal{C}_\tau+\eta_0.
\end{align}
Taking the time derivative of both sides of the above equation can yield the evolution equation of $\eta_t$,
\begin{align}\label{eq:diff-form-Boltzmann}
    U^*_t\partial_t\eta_t+&U^*_t\pounds_L\eta_t=U^*_t\mathcal{C}_t\quad\Rightarrow\quad\partial_t\eta_t+\pounds_L\eta_t=\mathcal{C}_t,
\end{align}
where $\pounds_L$ denotes the Lie derivative in the direction of $L$.

Let $\mathcal{N}$ be a hypersurface of $M$. The intersection points per unit time between the paths of these particles and $\mathcal{N}$ form a distribution, which is essentially a flux on $\mathcal{N}$. We will provide a theorem to show how $\eta_t$ relates to this distribution. Before stating this theorem, we introduce a lemma for convenience.

\begin{lemma}\label{lemma:coarea}
    If the hypersurface $\mathcal{V}_t$ of the manifold $M$ moves along $-L$, i.e.
    \[
    \mathcal{V}_t = U_t^{-1}[\mathcal{V}],
    \]
    then the integral of a full-rank form $\varepsilon$ over the area $\Gamma[t_1, t_2] := \bigcup_{\tau \in [t_1, t_2]} \mathcal{V}_\tau$ swept by $\mathcal{V}_t$ is given by
    \[
    \int_{\Gamma[t_1, t_2]} \varepsilon = \int_{t_1}^{t_2} \D\tau \int_{\mathcal{V}_\tau} \iota_L \varepsilon.
    \]

    \begin{proof}
        Extend the subscript of $\mathcal{V}_\tau$ to a scalar field $\hat{\tau}$ on $\Gamma[t_1, t_2]$ such that $\hat{\tau}(x) = -\tau$ for any $x \in \mathcal{V}_\tau$. Clearly, the Lie derivative of $\hat{\tau}$ along $L$ is 1, \ie
        \[
        L(\hat{\tau}) = \pounds_L \hat{\tau} = 1,
        \]
        since $U_{\D\tau}$ can be viewed as a diffeomorphism from $\mathcal{V}_\tau$ to $\mathcal{V}_{\tau - \D\tau}$. 
        The fact that $\varepsilon$ is a top form gives
        $\D\hat{\tau}\wedge\varepsilon=0$, and therefore
        \[
        0=\iota_L(\D\hat{\tau}\wedge\varepsilon)
        =L(\hat{\tau})\varepsilon-\D\hat{\tau}\wedge\iota_L\varepsilon.
        \]
        Using $L(\hat{\tau})=1$, we obtain
        \[
        \varepsilon=\D\hat{\tau}\wedge\iota_L\varepsilon.
        \]
        Since $\mathcal{V}_\tau$ can be viewed as the hypersurface where $\hat{\tau}$ takes a constant value, the integral of $\varepsilon$ over $\Gamma[t_1, t_2]$ can be written as
        \[
        \int_{\Gamma[t_1, t_2]} \varepsilon = \int_{\Gamma[t_1, t_2]} \D\hat{\tau} \wedge \iota_L \varepsilon = \int_{t_1}^{t_2} \D\tau \int_{\mathcal{V}_\tau} \iota_L \varepsilon.
        \]
    \end{proof}
\end{lemma}

\begin{theorem}\label{thm:flux}
    Let $\mathcal{N}$ be an arbitrary hypersurface within $M$. There exists a full-rank form $\xi_t$ on $\mathcal{N}$ such that the particle number $N$ through an area $\mathcal{V}$ of $\mathcal{N}$ from $t_1$ to $t_2$ reads
    \[
    N=\int_{t_1}^{t_2}\D\tau\int_{\mathcal{V}}\xi_\tau.
    \]
    $\xi_t$ is referred to as the flux in this paper. The flux associated with the distribution $\eta_t$, whose evolution is determined by eq.~\eqref{eq:diff-form-Boltzmann}, is given by
    \[
        \xi_t=\iota_L\eta_t.
    \]
    \begin{proof}
        Let us consider the problem from the opposite perspective: keep the particles stationary and let $\mathcal{V}_t := U^{-1}_t[\mathcal{V}]$ move in the direction of $-L$. The number of stationary particles swept by $\mathcal{V}_t$ equals the number of moving particles passing through $\mathcal{V}$. Although the particles are stationary, their distribution evolves due to the particle source $\mathcal{C}_t$. The time evolution of the distribution of stationary particles is given by
        \[
        \hat\eta_t = U^*_t \eta_t.
        \]
        According to Lem.~\ref{lemma:coarea}, the number of particles swept by $\mathcal{V}_t$ during an infinitesimal time $\D t$, is
        \[
        \D N = \D t \int_{\mathcal{V}_t} \iota_L \hat\eta_t.
        \]
        Since $U_{-t}$ can be considered a diffeomorphism from $\mathcal{V}$ to $\mathcal{V}_t$, the integral in the above equation can be rewritten as an integral over $\mathcal{V}$,
        \[
        \D N = \D t \int_{\mathcal{V}} U^*_{-t} \iota_L \hat\eta_t = \D t \int_{\mathcal{V}} \iota_L \eta_t,
        \]
        where $U^*_{-t} L = L$ is used. Therefore, the total number of particles passing through $\mathcal{V}$ from $t_1$ to $t_2$ is
        \[
        N = \int_{t_1}^{t_2} \D N = \int_{t_1}^{t_2} \D t \int_{\mathcal{V}} \iota_L \eta_t=\int_{t_1}^{t_2} \D t \int_{\mathcal{V}}\xi_t.
        \]
        Since $\mathcal{V}$ is arbitrary, the result follows.
    \end{proof}
\end{theorem}

\begin{theorem}\label{thm:flux-convergence}
    Let $\mathcal{N}$ be a hypersurface of $M$ transverse to $L$.
    Assume that each integral curve of $L$ intersects
    $\mathcal{N}$ at most once. For a full-rank form, let
    $\lvert\cdot\rvert$ denote its absolute density, and use the same
    convention for the flux on $\mathcal{N}$. Suppose that, for some
    $t_0\in\mathbb{R}$,
    \begin{align}\label{eq:source-integrability}
        \int_M\lvert\eta_{t_0}\rvert<\infty,
        \qquad
        \int_{\mathbb{R}}\D t\int_M\lvert\mathcal{C}_t\rvert<\infty.
    \end{align}
    Then the flux $\xi_t=\iota_L\eta_t$ is absolutely integrable over
    $\mathbb{R}\times\mathcal{N}$. In particular, the integrated flux
    \begin{align}
        \int_{\mathbb{R}}\D t\int_{\mathcal{N}}\xi_t
    \end{align}
    converges absolutely.

    \begin{proof}
        The distribution $\eta_t$ is governed by eq.~\eqref{eq:diff-form-Boltzmann}. Applying eq.~\eqref{eq:particle-number-Vt-2} with $t_0$ as the initial time gives
        \begin{align}\label{eq:flux-convergence-solution}
            \eta_t=U_{t_0-t}^*\eta_{t_0}
            +\int_{t_0}^{t}\D s\,U_{s-t}^*\mathcal{C}_s.
        \end{align}
        Since $\xi_t=\iota_L\eta_t$, the triangle inequality gives
        \begin{align}
            \int_{\mathbb{R}}\D t\int_{\mathcal{N}}\lvert\xi_t\rvert
            \leq\int_{\mathbb{R}}\D t\int_{\mathcal{N}}
            \lvert\iota_LU_{t_0-t}^*\eta_{t_0}\rvert+\int_{\mathbb{R}}\D t\int_{\mathcal{N}}
            \left\lvert\iota_L\int_{t_0}^{t}\D s\,
            U_{s-t}^*\mathcal{C}_s\right\rvert.
            \label{eq:flux-term-splitting}
        \end{align}
        We estimate the two terms on the right-hand side in the order
        in which they appear.

        For the initial-data term, the map $U_{t_0-t}$ sends
        $\mathcal{N}$ to $U_{t_0-t}[\mathcal{N}]$ and preserves $L$.
        Hence, for every $t$,
        \begin{align}
            \int_{\mathcal{N}}
            \lvert\iota_LU_{t_0-t}^*\eta_{t_0}\rvert
            =\int_{U_{t_0-t}[\mathcal{N}]}
            \lvert\iota_L\eta_{t_0}\rvert.
        \end{align}
        As $t$ varies, the hypersurfaces
        $U_{t_0-t}[\mathcal{N}]$ move along $-L$. The no-return
        assumption implies that they do not overlap, so their union is
        the region swept out by $\mathcal{N}$. Applying
        Lem.~\ref{lemma:coarea} to the absolute density
        $\lvert\eta_{t_0}\rvert$, first on bounded $t$-intervals and
        then in the limit, gives
        \begin{align}\label{eq:initial-flux-bound}
            \int_{\mathbb{R}}\D t\int_{\mathcal{N}}
            \lvert\iota_L U_{t_0-t}^*\eta_{t_0}\rvert
            =\int_{\mathbb{R}}\D t
            \int_{U_{t_0-t}[\mathcal{N}]}
            \lvert\iota_L\eta_{t_0}\rvert
            =\int_{\bigcup_{t\in\mathbb{R}}
            U_{t_0-t}[\mathcal{N}]}
            \lvert\eta_{t_0}\rvert
            \leq\int_M\lvert\eta_{t_0}\rvert.
        \end{align}

        We next estimate the source term in
        eq.~\eqref{eq:flux-term-splitting}. For every
        $x\in\mathcal{N}$ and $t\in\mathbb{R}$, reversing the limits
        when $t<t_0$ changes only the sign of the integral, which
        disappears under the absolute value. The triangle inequality
        can therefore be displayed separately as
        \begin{align}
            \left\lvert\iota_L\int_{t_0}^{t}\D s\,
            U_{s-t}^*\mathcal{C}_s\right\rvert_x
            =\left\lvert
            \int_{\min(t,t_0)}^{\max(t,t_0)}\D s\,
            \iota_LU_{s-t}^*\mathcal{C}_s\right\rvert_x
            \leq\int_{\min(t,t_0)}^{\max(t,t_0)}\D s\,
            \left\lvert\iota_LU_{s-t}^*\mathcal{C}_s\right\rvert_x.
        \end{align}
        Integrating this pointwise inequality over $\mathcal{N}$,
        exchanging the order of integration, and then extending the
        nonnegative $s$ integral to the whole real line gives
        \begin{align}
            \int_{\mathcal{N}}
            \left\lvert\iota_L\int_{t_0}^{t}\D s\,
            U_{s-t}^*\mathcal{C}_s\right\rvert\leq
            \int_{\min(t,t_0)}^{\max(t,t_0)}\D s
            \int_{\mathcal{N}}
            \lvert\iota_LU_{s-t}^*\mathcal{C}_s\rvert
            \leq\int_{\mathbb{R}}\D s\int_{\mathcal{N}}
            \lvert\iota_LU_{s-t}^*\mathcal{C}_s\rvert.
        \end{align}
        The integrand on the right-hand side is nonnegative. We may
        therefore integrate over $t$, exchange the $s$ and
        $t$ integrations, and obtain
        \begin{align}
            \int_{\mathbb{R}}\D t\int_{\mathcal{N}}
            \left\lvert\iota_L\int_{t_0}^{t}\D s\,
            U_{s-t}^*\mathcal{C}_s\right\rvert
            \leq\int_{\mathbb{R}}\D s\int_{\mathbb{R}}\D t
            \int_{\mathcal{N}}
            \lvert\iota_LU_{s-t}^*\mathcal{C}_s\rvert.
            \label{eq:source-flux-bound}
        \end{align}
        It remains to evaluate the inner $t$ integral for each fixed
        $s$. For every $t$, the map $U_{s-t}$ sends $\mathcal{N}$ to
        $U_{s-t}[\mathcal{N}]$ and preserves $L$; hence,
        \begin{align}
            \int_{\mathcal{N}}
            \lvert\iota_LU_{s-t}^*\mathcal{C}_s\rvert
            =\int_{U_{s-t}[\mathcal{N}]}
            \lvert\iota_L\mathcal{C}_s\rvert.
        \end{align}
        For fixed $s$, the hypersurfaces $U_{s-t}[\mathcal{N}]$ move
        along $-L$ as $t$ varies and sweep out the region
        $\bigcup_{t\in\mathbb{R}}U_{s-t}[\mathcal{N}]$. The no-return
        assumption makes these hypersurfaces disjoint. Applying
        Lem.~\ref{lemma:coarea} first on bounded $t$-intervals and then
        in the limit therefore gives
        \begin{align}
            \int_{\mathbb{R}}\D t\int_{\mathcal{N}}
            \lvert\iota_LU_{s-t}^*\mathcal{C}_s\rvert=\int_{\mathbb{R}}\D t
            \int_{U_{s-t}[\mathcal{N}]}
            \lvert\iota_L\mathcal{C}_s\rvert
            =\int_{\bigcup_{t\in\mathbb{R}}U_{s-t}[\mathcal{N}]}
            \lvert\mathcal{C}_s\rvert
            \leq\int_M\lvert\mathcal{C}_s\rvert.
        \end{align}
        Substituting this estimate into
        eq.~\eqref{eq:source-flux-bound}, and then using
        eqs.~\eqref{eq:flux-term-splitting} and
        \eqref{eq:initial-flux-bound}, we obtain
        \begin{align}
            \int_{\mathbb{R}}\D t\int_{\mathcal{N}}\lvert\xi_t\rvert
            \leq \int_M\lvert\eta_{t_0}\rvert
            +\int_{\mathbb{R}}\D s\int_M\lvert\mathcal{C}_s\rvert<\infty.
        \end{align}
        Thus the flux integral is absolutely convergent.
    \end{proof}
\end{theorem}

\begin{theorem}
    Let $V_n$ be a sequence of areas of $M$ such that $\lim_{n\rightarrow \infty}V_n=M$. If the limit of the integral of the flux $\xi_t$ over the boundary of $V_n$ is equal to zero, i.e.
    \[
    \lim_{n\rightarrow \infty}\int_{\partial V_n}\xi_t=0,
    \]
    the time derivative of the total particle number is the integral of the particle source $\mathcal{C}_t$ over $M$,
    \[
    \frac{\D }{\D t}\int_M\eta_t=\int_M \mathcal{C}_t.
    \]
    \begin{proof}
        Using the Cartan formula, it is easy to prove that the Lie derivative of the distribution $\eta_t$ in the direction of $L$ is equal to the exterior derivative of the flux $\xi_t$,
        \[
        \pounds_L\eta_t=\D\iota_L\eta_t=\D\xi_t.
        \] 
        Therefore, the evolution equation \eqref{eq:diff-form-Boltzmann} of the distribution $\eta_t$ can be rewritten as the form of continuity equation,
        \[
        \partial_t\eta_t+\D\xi_t=\mathcal{C}_t.
        \]
        Integrating both sides over $V_n$ and utilizing Stocks' theorem can yield
        \[
        \frac{\D }{\D t}\int_{V_n}\eta_t+\int_{\partial V_n}\xi_t=\int_{V_n}\mathcal{C}_t.
        \] 
        Let $n\rightarrow\infty$, the result follows.
    \end{proof}
\end{theorem}

\begin{lemma}\label{lem:restriction}
    Let $\mathcal{N}$ be a hypersurface within $ M $, $\varepsilon$ be a full-rank form on $M$ and $l$ be a vector tangent to $\mathcal{N}$, then the restriction of $\iota_l\varepsilon$ on $\mathcal{N}$ is zero.
    \begin{proof}
        Let $\{\mathcal{X}_1,\mathcal{X}_2,\cdots ,\mathcal{X}_{n-1}\}$ be an arbitrary set of vector tangent to $\mathcal{N}$, then they must be linearly dependent with $l$ since the dimension of $\mathcal{N}$ is $n-1$. It implies
        \[
            \iota_l\varepsilon(\mathcal{X}_1,\mathcal{X}_2,\cdots ,\mathcal{X}_{n-1})=\varepsilon(l,\mathcal{X}_1,\mathcal{X}_2,\cdots ,\mathcal{X}_{n-1})=0.
        \]        
        Therefore, the restriction of $\iota_l\varepsilon$ on $\mathcal{N}$ is zero.
    \end{proof}
\end{lemma}

\begin{corollary}\label{corc:Riemann-flux}
    \textup{\textbf{Scalarization}:}
    Let $ g $ be a Riemannian metric on $ M $, and $ \vol_{M} $ be the volume element associated with $ g $. Then the distribution $ \eta_t $ can always be written as a product between a scalar field $ f_t $ and $ \vol_{M} $,
    \begin{align}\label{eq:Riemann-flux-distribution-function}
        \eta_t = f_t \vol_M.
    \end{align}
    We refer to $ f_t $ as the distribution function under the volume element $ \vol_{M} $. Similarly, $ \mathcal{C}_t = C_t \vol_M $. Using the distribution function, we define the particle current as $ J := f_t L $. Consequently, the flux on $ \mathcal{N} $ can be written as
    \begin{align}
        \xi_t = g(J, n) \vol_{\mathcal{N}},
    \end{align}
    where $ n $ is the unit normal vector of $ \mathcal{N} $ and $ \vol_{\mathcal{N}} := \iota_n \vol_M $ is the volume element of $ \mathcal{N} $. The exterior derivative of the flux can be expressed as the product of the divergence of the particle current $ J $ and the volume element $ \vol_M $,
    \begin{align}\label{eq:Riemann-flux-exterior-derivative}
        \D \xi_t = (\mathrm{div} J) \vol_M.
    \end{align}
    Furthermore, if the time evolution is volume preserving, \ie, $ \pounds_L \vol_{M} = 0 $, the time evolution equation of $ f_t $ becomes
    \begin{align}\label{eq:Riemann-flux-evolution}
        \partial_t f_t + L(f_t) = C_t,
    \end{align}
    which can be rewritten in the form of the continuity equation,
    \begin{align}\label{eq:Riemann-flux-evolution2}
        \partial_t f_t + \mathrm{div} J = C_t.
    \end{align}
    Specifically, if $ M $ is the phase space of a single particle, $ L $ is the Liouville vector of a free particle, and $ C_t $ is a scattering integral, eq.~\eqref{eq:Riemann-flux-evolution} is the Boltzmann equation.
    \begin{proof}
        Decompose $ L $ into components orthogonal and tangent to the hypersurface $ \mathcal{N} $,
        \begin{align}\label{decompose-L}
            L = g(L, n) n + l.
        \end{align}
        It is straightforward to verify that $ g(l, n) = 0 $; hence, $l$ is tangent to $\mathcal{N}$. According to Lem.~\ref{lem:restriction}, $ \iota_l \eta_t $ restricted to $ \mathcal{N} $ must be zero. According to Thm.~\ref{thm:flux}, the flux can be expressed by the interior product $ \xi_t = \iota_L \eta_t $. Rewriting the flux using the decomposition of $ L $ and $ \eta_t $,
        \begin{align}
            \xi_t &= \iota_L \eta_t = (g(L, n) \iota_n + \iota_l) \eta_t \notag \\
            &= g(L, n) f_t \iota_n \vol_M = g(J, n) \vol_{\mathcal{N}}.
        \end{align}
        Given that the covariant divergence of a vector $ A $ on the Riemann manifold $ (M, g) $ satisfies \cite{lee2012smooth,lee2022manifolds}
        \begin{align}
            (\mathrm{div} A) \vol_M = \pounds_A \vol_M,
        \end{align}
        eq.~\eqref{eq:Riemann-flux-exterior-derivative} follows:
        \begin{align}
            (\mathrm{div} J) \vol_M = \pounds_J \vol_M = \D (\iota_J \vol_M) = \D (f_t \iota_L \vol_M) = \D \iota_L \eta_t = \D \xi_t.
        \end{align}
        Substituting eq.~\eqref{eq:Riemann-flux-distribution-function} into eq.~\eqref{eq:diff-form-Boltzmann},
        \begin{align}
            \partial_t f_t \vol_M + \pounds_L (\eta_t) &= \partial_t f_t \vol_M + \pounds_L (f_t) \vol_M + f_t \pounds_L (\vol_M) \notag \\
            &= [\partial_t f_t + L(f_t)] \vol_M = C_t \vol_M.
        \end{align}
        Here, the assumption $ \pounds_L \vol_M = 0 $ is used. Thus, eq.~\eqref{eq:Riemann-flux-evolution} follows from the above equation. This assumption implies the divergence of $ L $ is zero,
        \begin{align}
            (\mathrm{div} L) \vol_M = \pounds_L \vol_M = 0.
        \end{align}
        Hence, the divergence of the particle current is
        \begin{align}
            \mathrm{div} J = L(f_t) + f_t \cdot \mathrm{div} L = L(f_t).
        \end{align}
        Substituting this into eq.~\eqref{eq:Riemann-flux-evolution} yields eq.~\eqref{eq:Riemann-flux-evolution2}.
    \end{proof}
\end{corollary}

\section{decomposition of boundary volume elements}

\begin{lemma}\label{lem:decomposition}
    Let $S$ be a hypersurface in spacetime $\mathcal{M}$, and define
    \[
        \Sigma:=\pi^{-1}[S]=\bigcup_{x\in S}(\Gamma_0^+)_x,
    \]
    where $\pi:\Gamma^+_0\rightarrow\mathcal{M}$ is the projection map.
    Let $V=V^\mu\partial_{x^\mu}$ be a vector field on $\mathcal{M}$, and let $\mathcal{V}=V^\mu e_\mu$ denote its horizontal lift to $\Gamma_0^+$. Upon restricting both sides to $\Sigma$, the following equality of differential forms holds:
    \begin{align}
        \iota_{\mathcal{V}}\vol_{\Gamma_0^+}
        =(\iota_V \vol_{\mathcal M})\wedge\varpi.
    \end{align}
    \begin{proof}
        Using eq.~\eqref{eq:decompose-volume-gamma}, and suppressing the pullback $\pi^*$ as in that equation, we have
        \begin{align}
            \iota_{\mathcal{V}}\vol_{\Gamma_0^+}
            &=\iota_{\mathcal{V}}(\vol_{\mathcal M}\wedge\varpi) \notag\\
            &=\bigl(\iota_{\mathcal{V}}\vol_{\mathcal M}\bigr)\wedge\varpi
            +(-1)^{d+1}\vol_{\mathcal M}\wedge\iota_{\mathcal{V}}\varpi.
        \end{align}
        Since $S$ is a hypersurface in $\mathcal{M}$, the coordinates $x^\mu$ are not all independent on $S$. Locally, after a possible relabeling of the coordinates, the hypersurface $S$ can be written as
        \begin{align}
            x^0=F(x^1,\ldots,x^d).
        \end{align}
        Consequently, on $\Sigma$, we have
        \begin{align}
            \D x^0=\sum_{i=1}^{d}\frac{\partial F}{\partial x^i}\D x^i,
        \end{align}
        and hence,
        \begin{align}
            \left.\vol_{\mathcal M}\right|_{\Sigma}
            =\left.\left(\sqrt{g}\,\D x^0\wedge\cdots\wedge\D x^d\right)\right|_{\Sigma}=0.
        \end{align}
        Therefore, the second term vanishes upon restriction to $\Sigma$.

        For the first term, using
        \begin{align}
            e_\mu=\partial_{x^\mu}-\varGamma^\nu{}_{\alpha\mu}k^\alpha\partial_{k^\nu}
        \end{align}
        and $\iota_{\partial_{k^\mu}}\D x^\nu=0$, we obtain
        \begin{align}
            \iota_{\mathcal{V}}\vol_{\mathcal M}
            &=V^\mu\iota_{e_\mu}\vol_{\mathcal M}
            =V^\mu\iota_{\partial_{x^\mu}}\vol_{\mathcal M}
            =\iota_V\vol_{\mathcal M}.
        \end{align}
        Combining these observations and restricting to $\Sigma$ gives
        \begin{align}
            \iota_{\mathcal{V}}\vol_{\Gamma_0^+}
            =(\iota_V\vol_{\mathcal M})\wedge\varpi.
        \end{align}
    \end{proof}
\end{lemma}

\end{appendices}

\end{document}